\documentclass{PoS}

\usepackage{wrapfig}

\newcommand\beq{\begin{eqnarray}}
\newcommand\eeq{\end{eqnarray}}
\newcommand {\bmp}{\begin{minipage}}
\newcommand {\emp}{\end{minipage}}
\newcommand \reffig {Fig.~\ref}
\newcommand \refeq {Eq.~\ref}
\newcommand \reftab {Tab.~\ref}

\title{Extended study for unitary fermions on a lattice using the cumulant expansion technique}

\ShortTitle{Extended study for unitary fermions on a lattice using the cumulant expansion technique}

\author{\speaker{Jong-Wan Lee}\\
        Institute for Nuclear Theory, University of Washington, Seattle, WA 98195-1550, USA and\\
	KEK Theory Center, Institute of Particle and Nuclear Studies, 
	High Energy Accelerator Research Organization (KEK), Tsukuba 305-0801, Japan\\
        E-mail: \email{jongwan@post.kek.jp}}

\author{Michael G. Endres\\
		Theoretical Physics Laboratory, RIKEN, Wako, Saitama 351-0198, Japan\\
		E-mail: \email{endres@riken.jp}}

\author{David B. Kaplan\\
		Institute for Nuclear Theory, University of Washington, Seattle, WA 98195-1550, USA\\
		E-mail: \email{dbkaplan@washington.edu}}

\author{Amy N. Nicholson\\
		Institute for Nuclear Theory, University of Washington, Seattle, WA 98195-1550, USA and\\
		Deptartment of Physics, University of Maryland, College Park MD 20742-4111, USA\\
		E-mail: \email{amynn@umd.edu}}

\abstract{
A recently developed lattice method for large numbers of strongly interacting nonrelativistic fermions exhibits a heavy tail in the distributions 
of correlators for large Euclidean time $\tau$ and large number of fermions $N$, which only allows the measurement of ground state energies 
for a limited number of fermions using standard techniques. In such cases, it is suggested that measuring the log of the correlator is more efficient, 
and a cumulant expansion of this quantity can be exactly related to the correlation function. The cumulant expansion technique 
allows us to determine the ground state energies of up to $66$ unpolarized unitary fermions on lattices as large as $72 \times 14^3$, 
and up to $70$ unpolarized unitary fermions trapped in a harmonic potential on lattices as large as $72 \times 64^3$.
We have also improved our lattice action with a Galilean invariant form for the four-fermion interaction, 
which results in predictive volume scaling of the lowest energy of three fermions in a periodic box 
and in good agreement of our results for $N\leq 6$ trapped unitary fermions 
with those from other benchmark calculations.
}
\FullConference{The XXIX International Symposium on Lattice Field Theory - Lattice2011\\
		July 10-16, 2011\\
		Squaw Valley, Lake Tahoe, CA, USA}

\begin{document}

\section{Introduction}

One of the most challenging problems in physics is the quantitative understanding 
of a quantum system of many strongly interacting particles 
for which numerical simulations have played an important role. 
Among the variety of known systems, a dilute Fermi gas near unitarity has been noted 
for its pure form, which can be readily studied from theory, and for its realization by using 
ultra-cold atomic experiments (for a recent review, see \cite{Giorgini, Inguscio} ). 
Besides the intrinsic physical interest, such as universal behavior, 
the unitary Fermi gas might also be considered as an ideal starting point to develop numerical techniques 
which can be applied to low energy nuclear physics, and to attack the noise problem which 
typically appears in numerical simulations of many particles. 

In 2010 we presented a highly improved lattice method for non-relativistic fermions with four-fermion contact interactions \cite{Endres} 
and its applications for trapped and untrapped unitary fermions \cite{JW, Nicholson}, 
where the largest numbers of fermions were restricted to $N=20$ and $38$, respectively, due to a statistical overlap problem.  
Since then, we have devised a tuning technique with a Galilean invariant form for the four-fermion interaction, 
which keeps the system at unitary up to Galilean boosts. We have also 
implemented an external harmonic potential in a more sophisticated way. 
Finally, we have developed a cumulant expansion technique 
to extract the ground state energies from data exhibiting a distribution overlap problem \cite{Noise, Trapped}. 
As a result, we are able to extend our calculation of ground state energies of unpolarized fermions up to $N=70$ 
and up to $N=66$ unitary fermions with and without a harmonic trap, respectively.  
For the untrapped case, this extension leads us to the regime, $N\geq 38$, where the ground state energy $E$ in units of 
that for non-interacting particles $E^{(0)}$\footnote{
The energies of non-interacting untrapped and trapped fermions are given by 
$E_{\textrm{untrapped}}^{(0)}=(3N)^{(5/3)}\pi^{4/3}/10ML^2$ 
and $E_{\textrm{trapped}}^{(0)}=(3N)^{4/3}\omega/4$, respectively. Here, $\omega$ is an oscillator frequency.
} 
is constant; this implies that we are near the thermodynamic limit 
and can determine the universal dimensionless parameter $\xi=E/E^{(0)}$, 
called the Bertsch parameter. 
On the other hand, for the trapped case we find that $N=70$ is insufficient to reach the thermodynamic limit. 

\section{Lattice construction for fermions at unitarity}
\label{sec:lattice_construction}

We consider simulations of $N$ non-relativistic two-component fermions $\psi=(\psi^{\uparrow},\psi^{\downarrow})$  
on a $T\times L^3$ Euclidean lattice, where periodic boundary conditions 
on the spatial directions and an open boundary condition on the temporal direction are imposed. 
The four-fermion contact interaction between different species of fermions 
is generated by $Z_2$ auxiliary
fields $\phi$ which live on the time-like links of the lattice. 
A consequence of this set up is that the fermion-determinant does not depend on the $\phi$ field 
and thus the quenched simulation is exact \cite{Endres}. 

The discretization errors for the energy of a single fermion are eliminated through the use of a perfect dispersion relation, 
while the unitarity limit may be achieved by tuning the interaction strength $C({\bf q})$ to reproduce scattering data, 
$p\cot\delta_0=0$, using the L\"uscher formula \cite{Endres, Trapped}. 
Here the operator ${\bf q}$ for $|{\bf q}|<\Lambda$ corresponds to 
the momentum transfer between incoming and outgoing fermions, which makes the interaction Galilean invariant. 
The periodic function $C({\bf q})$ is defined by
\beq
C({\bf q})=\frac{4\pi}{M}\sum_{n=0}^{N_{O}-1} C_{2n} O_{2n}({\bf q}),
\eeq
where the basis functions are
\begin{eqnarray}
O_{2n}({\bf q}) = M_0^n\times 
\left\{ 
\begin{array}{ll}
\left(1-e^{-{\bf q}^2/M_0}\right)^n & |{\bf q}|\leq\Lambda \\  
\left(1-e^{-\Lambda^2/M_0}\right)^n  & |{\bf q}|>\Lambda 
\end{array}
, \right.
\end{eqnarray}
for ${\bf q}$ within the first Brillouin zone and periodic from one Brillouin zone to the next. 
Although the parameter $M_0$ may in general be different from the fermion mass $M$, we use $M_0=M$ in this work. 

\begin{figure}
\bmp[t]{.48\linewidth}
\centering
\includegraphics[width=1.0\textwidth]{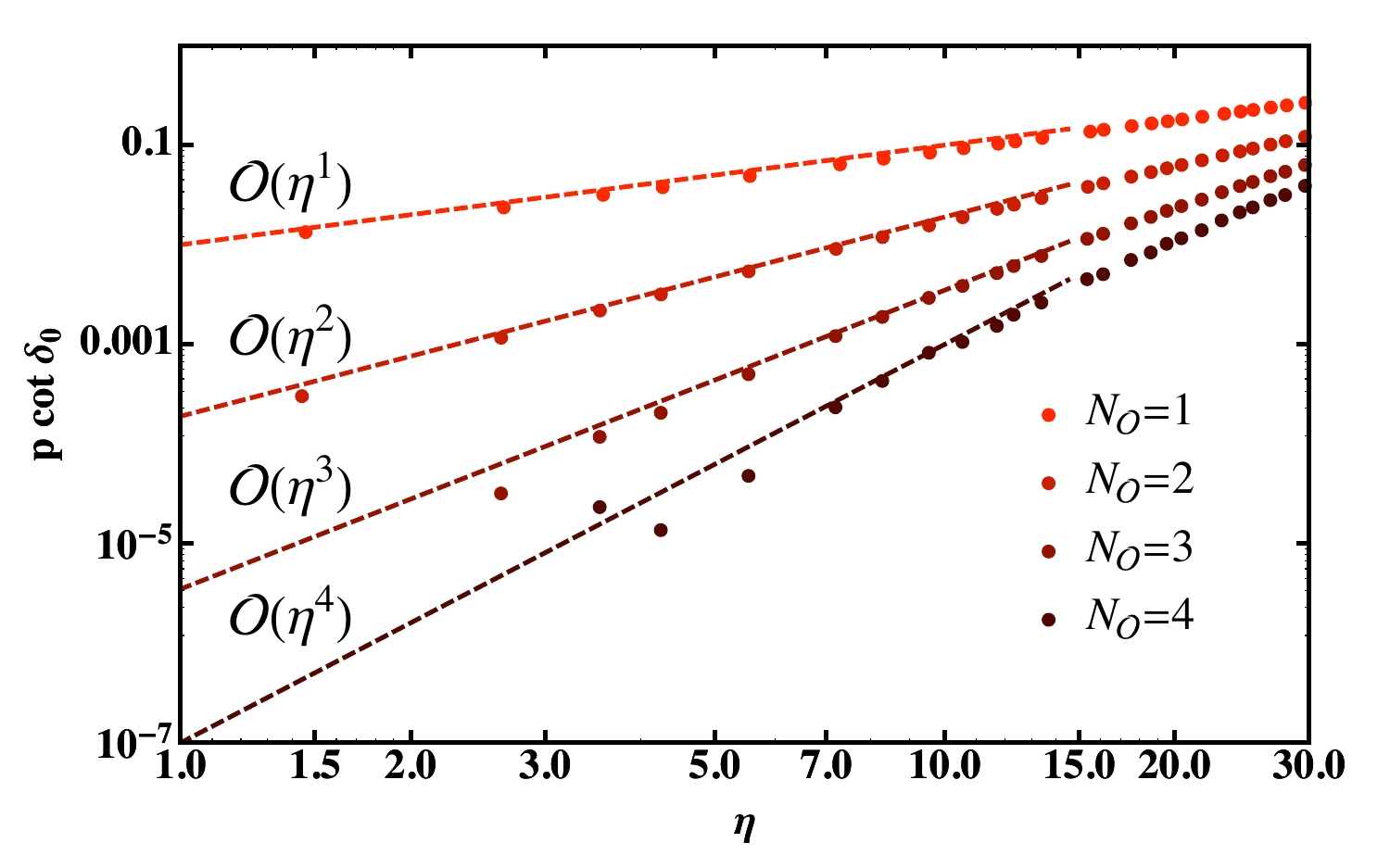}
\caption{ln-ln plot of $p\cot\delta_0$ (dots) along with expected $\eta$ scaling (dashed lines) 
for which $N_O=1,2,3$ and $4$ coefficients are tuned by exactly matching the first $N_O$ energy eigenvalues 
for two particles in a box. 
Data is from an $L=32$ and $M=5$ lattice.}\label{fig1}
\emp
\hskip .2in
\bmp[t]{.48\linewidth}
\includegraphics[width=1.0\textwidth]{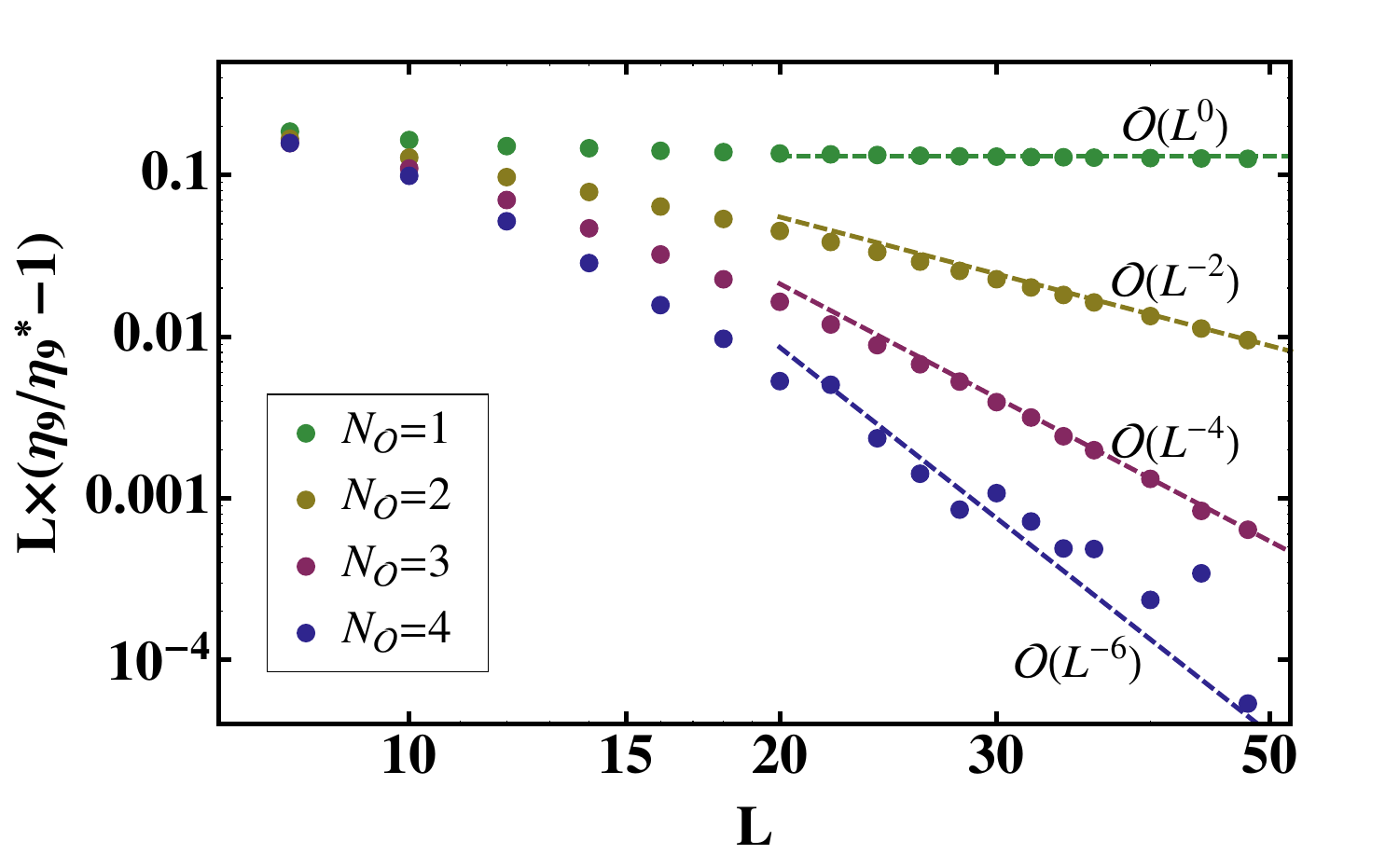}
\caption{
$L$ dependence of the $9$th energy eigenvalues for two particles in a box. 
The dashed lines are fit results of the data (dots) using Eq. 2.3.
}
\label{fig2}
\emp
\end{figure}

The coefficients $C_{2n}$ are numerically determined by matching the lowest $N_{O}$ energy eigenvalues of the 
two-body transfer matrix in our lattice theory 
with the lowest $N_{O}$ L\"uscher 
energy eigenvalues in the corresponding continuum theory. 
In \reffig{fig1} we plot the $p\cot\delta_0$ (dots), computed using the exact lattice energy eigenvalues and L\"uscher's formula, 
which results from tuning the first $N_O$ terms in the effective range expansion to zero. 
These results show an expected $\eta$ scaling represented by the dashed lines in the figure. 
A systematic improvement is also seen for the energies of excited states above 
the lowest $N_{O}$ tuned states, where the correction may be predicted by \cite{Trapped},
\beq
L\left(\frac{\eta_k}{\eta_k^*}-1\right)\propto L^{2-2n}.
\label{energy_eta_scaling}
\eeq
Here $\eta_k$ are the $k$th eigenvalues of the two-body transfer matrix with $N_O$ terms tuned, 
while $\eta_k^*$ are the $k$th solutions of L\"uscher's formula in the unitary limit.  
As an example, we plot $L(\eta_k/\eta_k^*-1)$ with respect to $L$ (dots) in \reffig{fig2} 
and find good agreement with \refeq{energy_eta_scaling}. 

In simulations of trapped unitary fermions, 
the external harmonic potential, $U=\frac{1}{2}\kappa {\bf x}^2$ with spring constant $\kappa$, 
has been implemented in the transfer matrix as
\beq
T_{trapped}&=&e^{-b_\tau {\bf p}^2/4M} e^{-b_\tau U} (1-b_\tau V) 
e^{-b_\tau U} e^{-b_\tau {\bf p}^2/4M} \nonumber \\
&=&e^{-b_\tau (\mathcal{H}+U)+O(b_\tau^3)},
\eeq
where $V$ represents the interaction and $\mathcal{H}+U$ is the target Hamiltonian for trapped unitary fermions. 
As seen in this formula, temporal discrerization errors appear at $O(b_\tau^2)$.\footnote{
Throughout this paper, $b_s$ and $b_\tau$ represent the spatial and temporal lattice spacings, respectively. 
} 

\section{Measurement, overlap problem, and cumulant expansion method}

The $N$-body correlators are constructed by taking determinants of the Slater matrix,
where each element is obtained by evolving an initial state (source) at $\tau=0$ 
with a single particle propagator and projecting onto a final state (sink) at $\tau=T$.
The choice of sources and sinks for untrapped and trapped unitary fermions, 
which gives superior wave function overlap by imposing two-particle correlations at the sinks, 
has been described in \cite{JW} and \cite{Nicholson}, respectively. 

One of the greatest difficulties in extending our study to large $N$ was the 
apparent upward drift of the effective mass plot at large Euclidean time. 
By investigating the distribution of $N$-body correlators, we found that this difficulty arises from 
a heavily long-tailed distribution, 
requiring an exponentially large number of samples before the central limit theorem becomes applicable.
In other words, the path-integral probability measure has small overlap with 
the dominant part of the operator being estimated.
On the other hand, the distribution of the log of correlators is nearly Gaussian, 
implying that the standard estimation of the lower moments should succeed with moderately sized ensembles. 
A general relation between the expectation value of $\mathcal{C}$ and that of $\ln \mathcal{C}$ is given by \cite{Noise}
\beq
\ln \langle \mathcal{C}(\tau) \rangle = \sum_{n=1}^\infty \frac{\kappa_n}{n!},
\label{cum_exp}
\eeq
where $\kappa_k$ is the $k$th cumulant of $\ln \mathcal{C}$. The generalized effective mass 
and the ground state energy associated with 
each partial sum in \refeq{cum_exp} may be written as \cite{Trapped}
\beq
m_{\textrm{eff}}^{N_k}(\tau)=-\frac{1}{\Delta\tau}
\sum_{n=1}^{N_k}\frac{1}{n!}\left[\kappa_n(\tau+\Delta\tau)-\kappa_n(\tau)
\right] ~~~\textrm{and}~~~
E_{N_k}=\lim_{\tau\rightarrow\infty}m_{\textrm{eff}}^{N_k}(\tau).
\label{gen_eff_mass}
\eeq
Since the statistical uncertainties typically increase as $N_k$ increases, 
one may determine the ideal value $N_k^*$ 
for which the statistical uncertainties and truncation errors are comparable.

\section{Ground state energy of unitary fermions}
To extract the energies of the system, we perform standard bootstrap resamplings 
and correlated $\chi^2$ fits to the plateau region of the generalized effective mass.
Fitting systematic errors are obtained by varing the end points of the fitting interval.
The quoted errors represent the combination of the statistical and fitting systematic errors added in quadrature. 
\subsection{Untrapped unitary fermions}

\begin{figure}
\bmp[t]{.48\linewidth}
\centering
\includegraphics[width=1.0\textwidth]{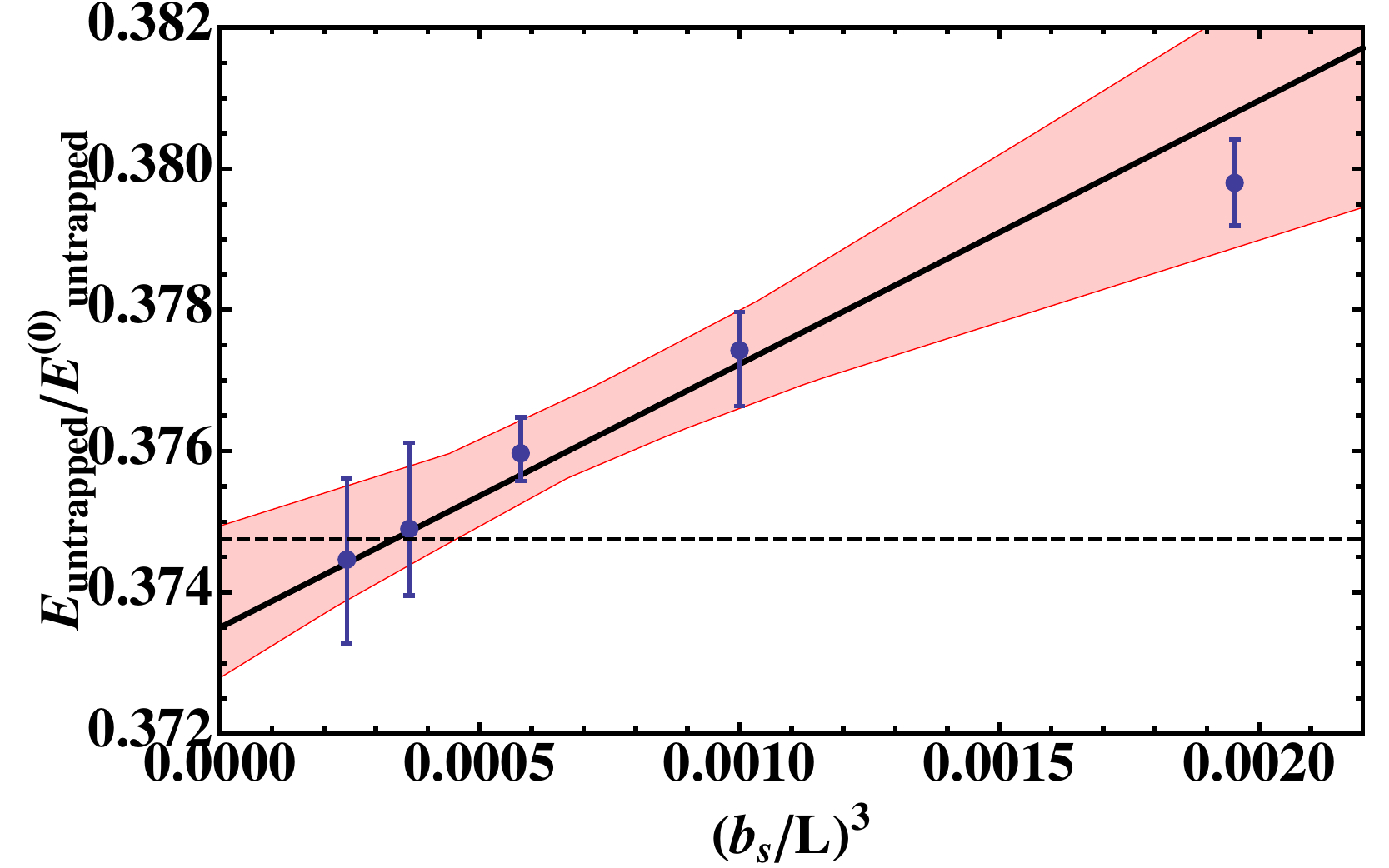}
\caption{
Energy of three untrapped unitary fermions in a zero total momentum eigenstate 
plotted versus $(b_s/L)^3$ for $L/b_s=8,10,12,14,16$. 
The red band represents the uncertainty in two-parameter fits of the $L/b_s\geq10$ data 
to the function $c_1+c_2/L^3$. 
while the black line is the fit to the central values. 
The dashed line is the result from Ref. \cite{Castin}. 
} \label{fig3}
\emp
\hskip .2in
\bmp[t]{.48\linewidth}
\includegraphics[width=1.0\textwidth]{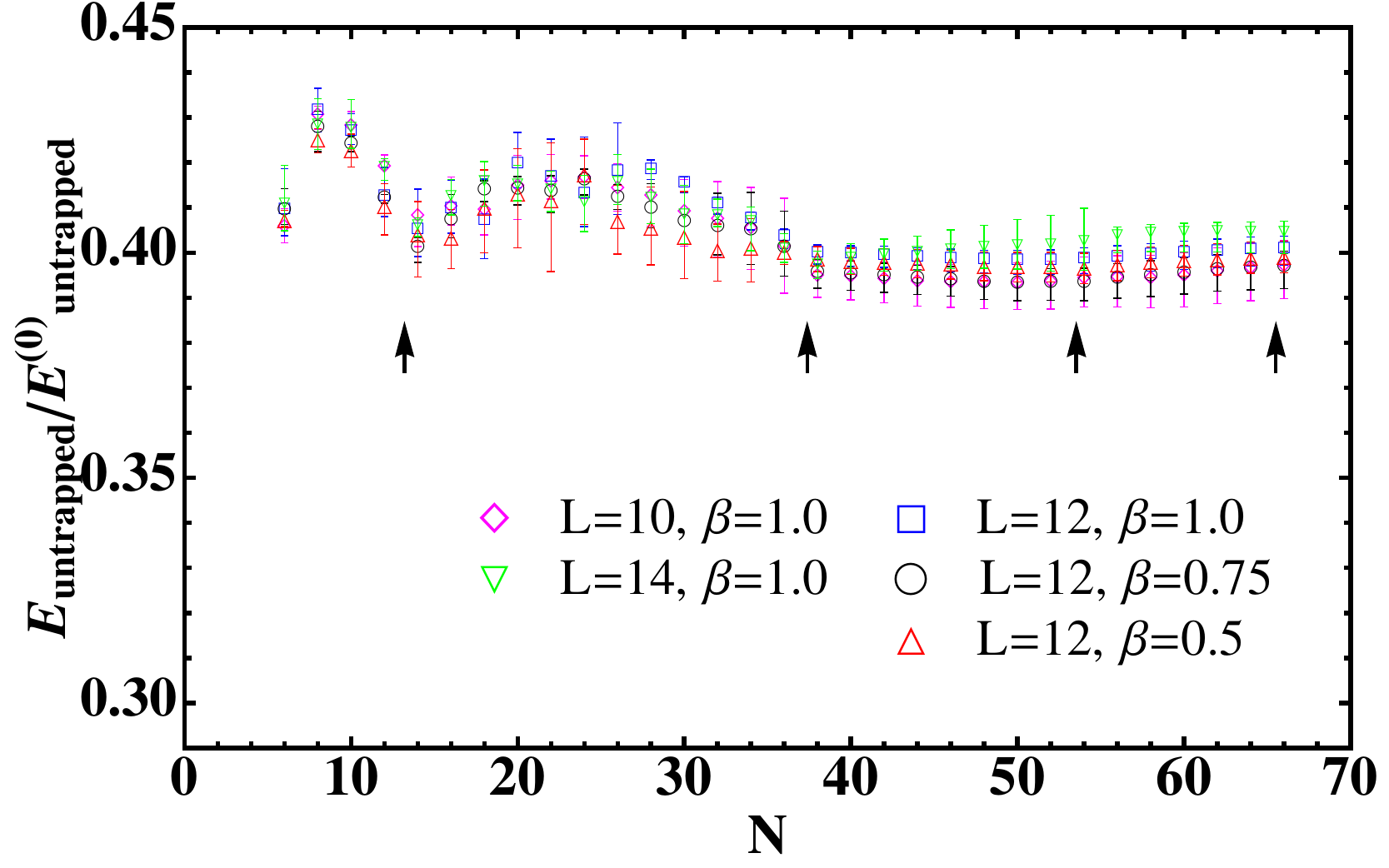}
\caption{
(Preliminary) Plot of ground state energies versus numbers of untrapped unitary fermions 
for $L=10, 12$ and $14$. 
For $L=12$ we considered three different sinks, $\beta=0.5, 0.75, 1.0$, to take into account 
any source dependence in our extraction of the energies. 
The arrows indicate the positions for which noninteracting fermions have fully occupied a given momentum shell.
} \label{fig4}
\emp
\end{figure}

As a nontrivial test of our lattice method, we have computed the lowest energy of 
three unitary fermions in a zero total momentum state with high precision. 
We performed the calculation for lattice sizes $L=8, 10, 12, 14, 16$, tuning the coefficients 
of four $O_{2n}$ operators for the $L=8$ lattice, and five for the other lattices. 
Since the single particle and two particle $s$-wave sectors have been highly tuned,
the leading $L$ dependence will be $L^{-3}$ due to the untuned two-derivative two-body 
$p$-wave operator. In \reffig{fig3} we plot the lowest energies versus $L^{-3}$. 
We perform two parameter fits of the $L\geq10$ data to $c_1+c_2 /L^3$ and find an infinite 
volume energy of $0.3735^{+0.0014}_{-0.0007}$ in units of the energy 
of three noninteracting fermions, which agrees with the high precision calculation 
by Pricoupenko and Castin \cite{Castin} 
within our $\sim 0.3\%$ uncertainty.

The preliminary results for the ground state energies of up to $66$ unpolarized unitary fermions 
in a periodic box are shown in \reffig{fig4}. 
With the given statistics, we do not resolve any shell structure in the energy 
for $N\geq 38$ and believe that the system is close to the thermodynamic limit. 
By averaging the results obtained from 
a constant fit of $\xi(N)$ for $N\geq 38$ using five different ensembles, 
we find a preliminary result for the Bertsch parameter, $\xi=0.399\pm 0.002$. 
The unitary Fermi gas has been extensively studied using the Quantum Monte Carlo (QMC) technique and 
the most recent calculation
reported is $\xi\leq0.383 (3)$ \cite{qmc_recent_1,qmc_recent_2}.
Recent experimental results for this parameter are $\xi=0.39 (2)$ \cite{Duke} 
and $0.41 (1)$ \cite{Paris}.


\subsection{Trapped unitary fermions}

\begin{figure}
\bmp[t]{.48\linewidth}
\centering
\includegraphics[width=1.0\textwidth]{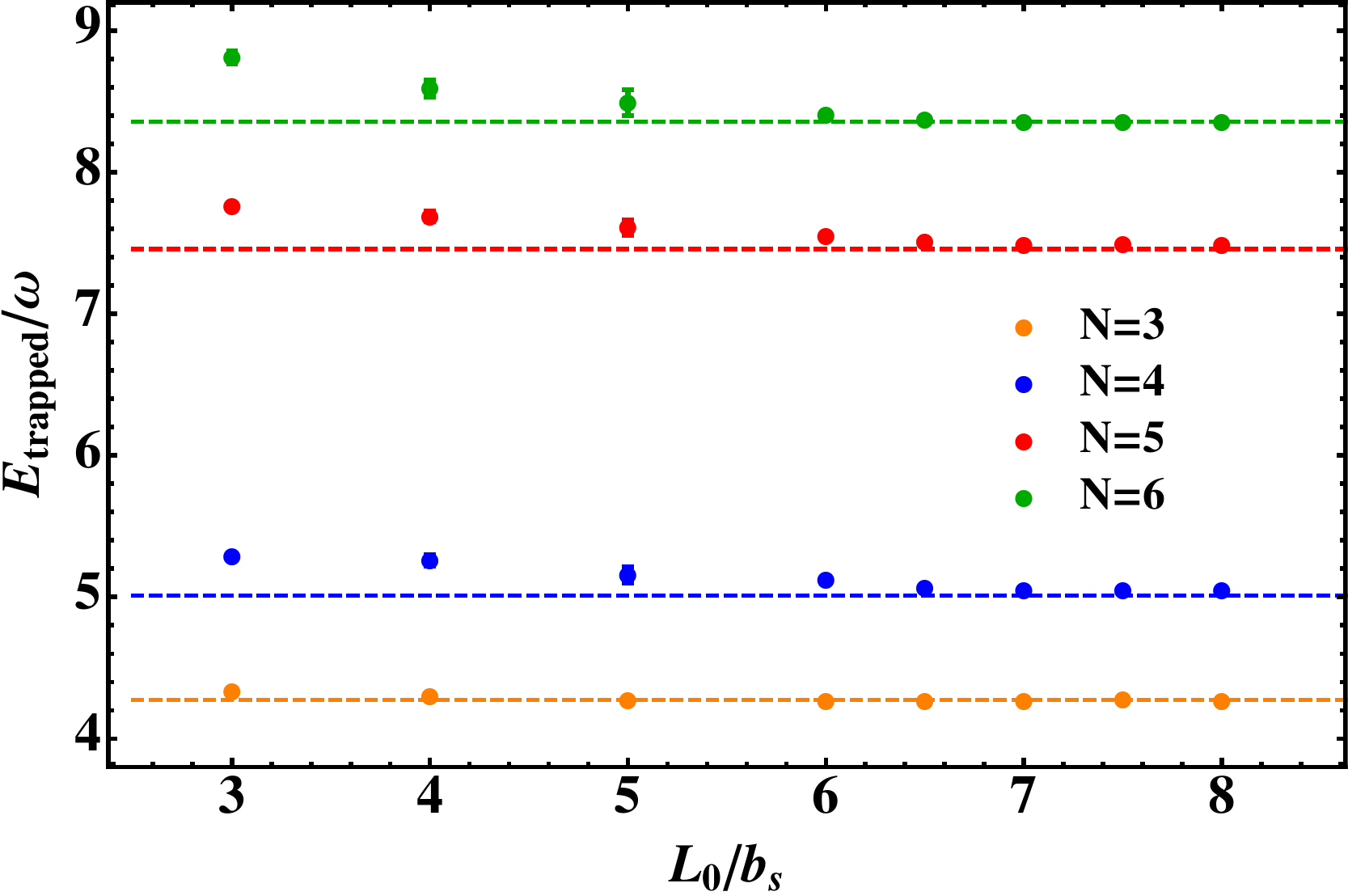}
\caption{
Ground-state energies (in units of $\omega$) as a function of $L_0/b_s$ at fixed $L/b_s = 48$ for various values of N. 
Dashed lines are results from Ref. \cite{Blume}.
}\label{fig5}
\emp
\hskip .2in
\bmp[t]{.48\linewidth}
\centering
\includegraphics[width=1.0\textwidth]{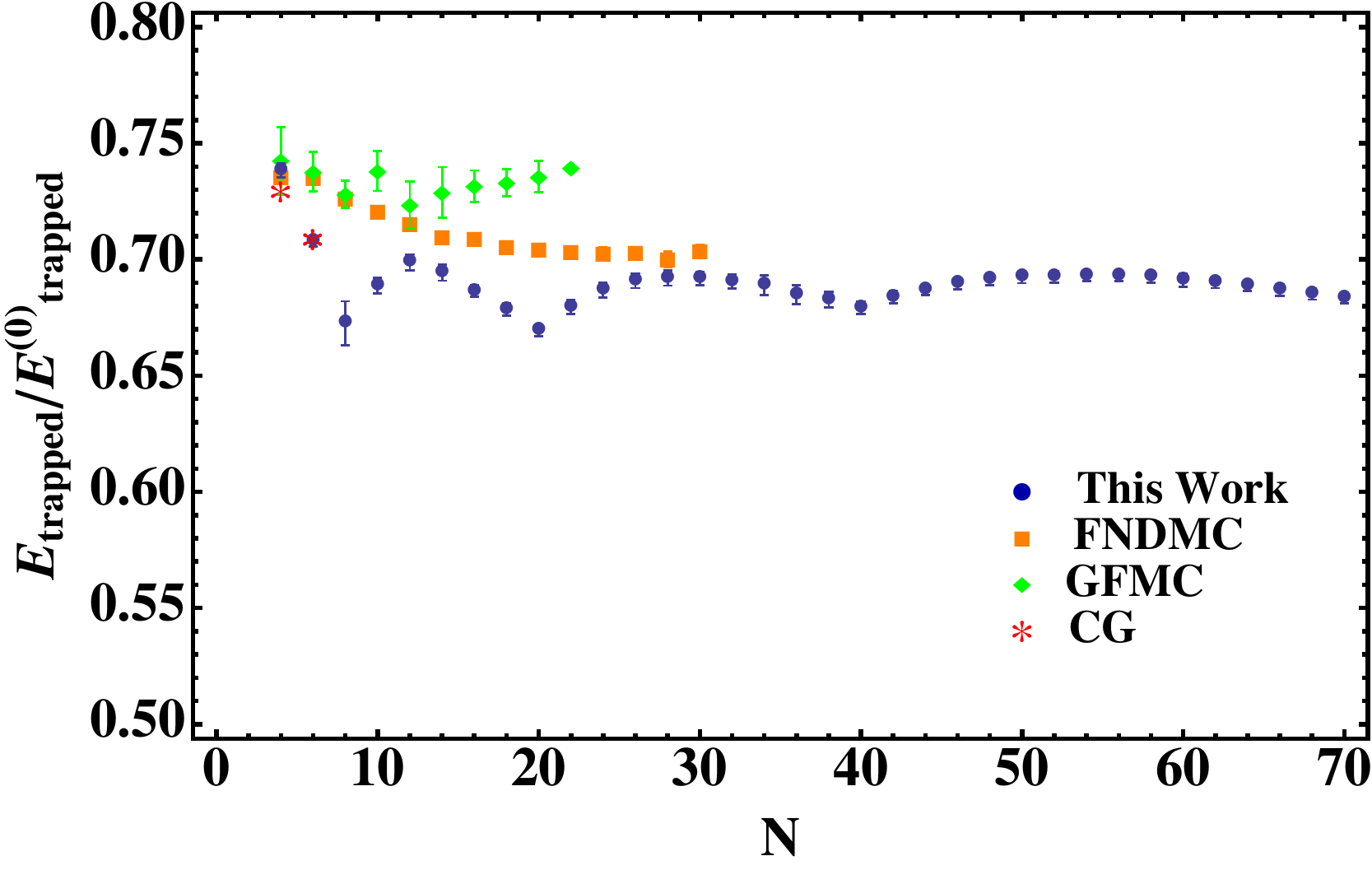}
\caption{
Ground-state energies of $N$-trapped unitary fermions. 
} \label{fig6}
\emp
\end{figure}

In numerical simulations of the trapped unitary Fermi gas, we have additional time scale $1/\omega$ 
and length scale $L_0=(M\omega)^{-1/2}$ which should be chosen 
so that $b_s\ll L_0 \ll L$ and $b_\tau \ll 1/\omega$ 
in order to minimize the discretization and finite volume errors. 
To balance the need for small temporal discretization errors with the computational cost 
associated with the number of time steps required to reach the ground state, 
we have chosen $\omega b_\tau=0.005$.
An ideal $L_0$ has been determined by scanning the parameter space of $L_0/b_s$ and $L/L_0$ 
and finding the region where both finite volume and spatial discretization errors are small. 
\reffig{fig5} presents our findings for the ground state energies of $N\leq 6$ fermions, 
with $L_0/b_s$ ranging from $3$ to $8$ and fixed $L/b_s=48$. 
For $L_0/b_s\leq7$ we find that the systematic errors increase as $L_0$ decreases, 
which indicates that the discretization error is not negligible. 
For $L_0/b_s\geq7$ we have also performed simulations with $L/b_s=64$, which showed no 
volume dependence; and thus we conclude that both types of systematic errors are negligible.   
For $6<N\leq70$, based on the results of the $L_0$ scan for $N\leq 6$, we have chosen to perform the calculation 
for three volumes ($L=48,54,64$) and at two values of the trap size ($L_0=7.5, 8$). 
We find non-negligible volume dependence and perform an infinite volume extrapolation 
for each $L_0$. 
All of these considerations are included along with statistical, fitting systematic, 
and truncation of the cumulant expansion errors in the final quoted errors of ground state energies. 

We first benchmark our method for up to $N=6$ against high-precision solutions 
to the many-body Schr\"odinger equation \cite{Blume}, 
achieving agreement at $1\%$ as shown in \reftab{trapped_few_body}. 
In \reffig{fig6} we plot the results of the ground state energies for $N\leq 70$ along with 
the results from 
two fixed-node calculations for comparison: a Green's function Monte Carlo (GFMC) approach \cite{GFMC} and 
a diffusion Monte Carlo (FN-DMC) approach \cite{DMC}, which provide upper bounds on the ground state energies. 
We find that our energies are consistently lower than those obtained using both of these methods. 
However, our results show clear shell structure which indicates that 
$N\sim 70$ is insufficient to reach the thermodynamic limit. 

\begin{table}
\caption{%
\label{tab:SmallN}%
Results for $E_\textrm{trapped}/\omega$ for $N \leq 6$, including combined statistical and fitting systematic errors (first row). 
For comparison we give the exact $N=3$ result \cite{Tan} and results of Ref. \cite{Blume} (second and third rows).}
\begin{center}
\begin{tabular}{|c|c|c|c|c|}
\hline
 & 3  & 4 & 5 & 6\\
\hline
this work & $4.243^{+0.037}_{-0.034} $ & $5.071^{+0.032}_{-0.075} $ & $7.511^{+0.051}_{-0.091} $ & $8.339^{+0.080}_{-0.066} $  \\
exact, Ref. \cite{Tan}  & 4.2727 & - & - & - \\
from Ref. \cite{Blume}  & 4.273(2)  &  5.008(1) & 7.458(10) & 8.358(20) \\
\hline
\end{tabular}
\end{center}
\label{trapped_few_body}
\end{table}


\section{Conclusion}

A new statistical technique, called the cumulant expansion method, allows us 
to extend our studies up to $N\sim 70$ unitary fermions 
both with and without an external harmonic trap. We report a preliminary value of $\xi=0.399\pm0.002$ 
from the calculation of the ground state energy for $N\geq 38$ untrapped unitary fermions. 
For $N\leq 70$ trapped fermions, we find that the shell structure is pronounced, which implies that 
we have not reached the thermodynamic limit, but our values of the ground state energies are consistently lower than 
those obtained from variational QMC calculations. 
In order to check the validity of our lattice method, we have performed high-precision calculations for $N=3$ 
untrapped fermions and $N\leq 6$ trapped fermions, and found good agreement with the results in \cite{Castin} and 
\cite{Tan, Blume}, respectively. 

In the near future, we will present results for the pairing gap by calculating the ground state energies of 
the slightly unpolarized unitary Fermi gas ($N_\downarrow=N_\uparrow+1$) and the integrated contact density 
by studying the dependence of the ground state energies on the $s$-wave scattering length near the unitary regime. 
Our lattice method combined with the new statistical technique is applicable for a wide range of 
nonrelativistic many-body systems.

\section{Acknowledgement}
This work was supported by U. S. Department of Energy grants DE-FG02-92ER40699 (M.G.E.) and DE-FG02-00ER41132 (D.B.K., J-W. L. and A.N.N.).
M.G.E is supported by the Foreign Postdoctoral Researcher program at RIKEN.
This research utilized resources at the New York Center for Computational Sciences at Stony Brook University/Brookhaven National Laboratory 
supported by the U.S. Department of Energy under Contract No. DE- AC02-98CH10886 and by the State of New York.
Computations for this work were also carried out in part on facilities of the USQCD Collaboration, 
which are funded by the Office of Science of the U.S. Department of Energy.

\end{document}